# Digital Storytelling for Competence Development in Games

Edgar Santos[1], Claudia Ribeiro[1], Manuel Fradinho[2], João Pereira[1],

[1] INESC-ID, Rua Alves Redol, Lisboa, Portugal
[2] Cyntelix, Business Inovation Center, Upper Newcastle Road, Galway, Irland
{edgar.santos, claudia.sofia.ribeiro}@ist.utl.pt;
mfradinho@cyntelix.com; jap@inesc.pt

**Abstract.** The acquisition of complex knowledge and competences raises difficult challenges for the supporting tools within the corporate environment, which digital storytelling presents a potential solution. Traditionally, a driving goal of digital storytelling is the generation of dramatic stories with human significance, but for learning purposes, the need for drama is complemented by the requirement of achieving particular learning outcomes. This paper presents a narrative engine that supports emergent storytelling to support the development of complex competences in the learning domains of project management and innovation. The approach is based on the adaptation on the Fabula model combined with cases representing situated contexts associated to particular competences. These cases are then triggered to influence the unfolding of the story such that a learner encounters dramatic points in the narrative where the associated competences need to be used. In addition to the description of the approach and corresponding narrative engine, an illustration is presented of how the competence "conflict management" influences a story

## 1 Introduction

Technology has fostered the collapse of the world into a small village, where businesses have become globally driven at an exceedingly fast pace to outperform their competitors to secure a share in a finite market. Human capital is recognized as key asset in an organization to achieve its goals, but the business, technology and societal pressures make corporate training a major concern where rapid competence development is the new mantra driving competitive advantage. However, the increase of complexity of knowledge and the constraint of acquiring it in shorter time frames imposes hard challenges to the competence development within the corporate environment. This is the case of soft skills, even more so when considering the richness of situations where one encounters multiple cultures. In addition, the digital native generation[1] have different expectations and learning patterns that compound the challenges of traditional approaches to corporate training.

The advent of serious games has given rise to the possibility of enhancing learning[2, 23] with an increasing number of advocates promoting the use of serious games as a delivery platform[3,4] for education and competence development. The benefits of serious games over more traditional learning methods and on-the-job training include the increase of learner motivation, ego gratification, fostering of creativity, socialization and above all making the experience fun. Although the body of evidence remains scarce, there is a growing evidence for the efficacy of serious games as educational tools with a growing number of research studies finding improved rates of learning and retention over more traditional learning methods (Druckman and Bjork, 1991; Charles and McAlister, 2004).

Since people are social beings that communicate and socialize with one another, the use of storytelling has played a crucial role in exchanging and transferring complex knowledge and foster understanding[5]. Therefore, it has been recognized that narratives are a valid support for learning because it helps make sense of experience and organize knowledge, in addition to increasing motivation (Dettori and Paiva). Therefore, it would beneficial to combine storytelling with serious games, but considering that a story implies the reflection of past events that unfolded with a particular order and that one of the strengths of serious games is the sense of agency of the learner[6], this raises a challenge known as Narrative Paradox (Aylett 2000). A possible approach to address this paradox is by allowing for "emergent narrative" (Ayllet, 1999), where the story emerges from characters' local interactions. Louchart et al. researched how an emergent narrative can be authored since it reflects the obvious paradox of the narrative being "authored" at run-time as it emerges from the characters' interactions. It usually implies an iterative process where the characters are modeled carefully, requiring several iterations are made to assure that the desired stories emerge and undesired stories are kept from happening. As stated by Crawford:

*"while architectural valid stories can be created by algorithm, humanly interesting stories can be created only by artists"*

Therefore, the authoring process can be costly and impose constraints on creation of viable stories that address desired learning outcomes. This paper presents an innovative approach to storytelling for the purpose of development of complex competences within the context of the TARGET project, which aims to support rapid competence development in project management and innovation. Consequently, the focus of the paper is the component – the Narrative Engine - responsible for the unfolding of an interactive story taking into account the learning outcomes in developing a particular set of competences along the lines of "sense-making", "conflict management", "time management".

## 2  Challenge

To better understand the challenge tackled by the TARGET project, one might begin by understanding the complexity of emergent storytelling. For such, one may consider the use of the story landscape[7] to demonstrate the space of possible story outcomes that can emerge from the underlying narrative engine. As the story unfolds, a plot across the landscape is made, which takes into consideration the surrounding hills and valleys.

An elevation in the landscape represent potential dramatic points in the story, which the narrative engine (or drama manager) will influence the story for a user to experience whilst a valley represents situations where a user is faced with multiple paths, but once climbing a hill, the dramatic context intensifies and moves in the particular direction of the elevation peak.

In the context of TARGET, the aim is not only to create engaging stories with emergent storytelling, but to promote particular learning outcomes, which leads to reshaping the landscape as the points are influenced not only by the dramatic element of a particular situation, but also its relevance to the learning plan of an individual.

## 3  Our Approach

As commonly referred when addressing the emergent narrative concept, stories will "emerge" from characters' local interactions. However, it is not enough since it lacks reasoning about pursuing story-level goals. The Narrative Engine (or more precisely the "Drama Manager") itself should be able to guarantee that an *interesting* and *learning-goal oriented* sequence of events is emerging. To do this kind of analysis, it is useful to build a formal representation of the emerging events.

### 3.1  Fabula Model

From a Narratology perspective, a distinction is often made between the *fabula* of the story and the *story*. *Fabula* is a series of causally and chronologically related events that are caused or experienced by characters in a story world. *Story* is a *fabula* that is looked at from a particular viewpoint. Considering these definitions, the Narrative Engine should document the sequence of events through a fabula, since it reflects an omniscient form of the narrative. Swartjes used such a representation in "The Virtual Storyteller"[10] and named it "Fabula Model", presented in Figure 2.

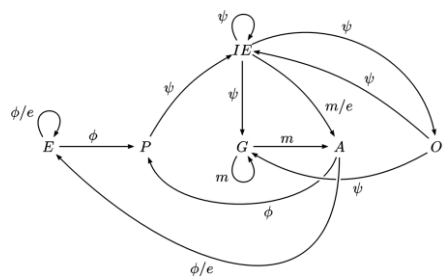

**Figure 2 - Fabula Model[11]**

After a critical analysis of the model it was noted/realized that the expressive power is due not only to the different elements (**E**vent, **P**erception, **I**nternal **E**lement, **G**oal,

**A**ction and **O**utcome), which are causally connected, but also because the causal connections have semantic meaning (ϕ – physical; ψ: psychological; *e*: enablement; *m*: motivational). These are necessary having into account the goal of the project in which they were used. In the Virtual Storyteller, a Plot Agent component uses this model, instantiates it with the events of the world simulation and a Narrator component can translate them to natural language. Writing a story in natural language takes a lot of semantic knowledge and it is understandable why the model needs to be so rich.

In TARGET, however, there is a need for formalizing story events and analyze them from a "narrative control" perspective and not from a semantic point of view. This concern represents the main motivation for adapting the model as explain in the next section.

### 3.2 Fabula Model Adaptation

In Figure 3, the proposed model for TARGET's fabula representation is presented.

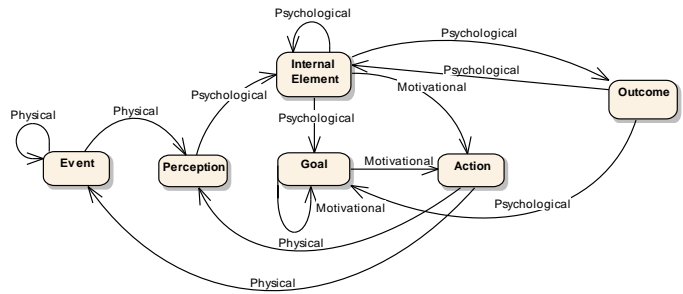

**Figure 3 - TARGET's Fabula Model**

According to the original model, "enablement" is the weakest form of causality. It represents the fact that "if a fabula element A enables another fabula element B, then B is possible because of A and no more than that"[11]. In practice it means that effects of A satisfy preconditions of B.

Since TARGET's purpose isn't to write a fabula in natural language, this kind of relation was discarded. We can implicitly think about their existence instead of making explicit such thinking. In fact, other types of causality exist only to assist the author in having a better mind model on how to build a narrative case (narrative cases will be covered in 3.3). Therefore, TARGET's fabula elements are causally connected but it is not necessary to make explicit the fact that those connections can be semantically different.

Other major difference is the use of the model with human characters. In "The Virtual Storyteller" human interaction is not taken into account. Internal Elements, Goals, Outcomes and event Perceptions of a player cannot be perceived by the system. A character controlled by a player is what we call "viewpoint character" (VC) and we only represent its' actions in the emerging fabula. The VC is a character through which a player "percepts" the emerging events into a story in a process called "storification"[12]. Notice that causality between VC's actions and their consequences still remains. What we don't represent is what caused a given VC's action. This is not a limitation at all since we don't want to control player's actions and we assume that the player will act "in-character" interpreting a given role as in the premises of Barros et al [13].

We still need a way to use fabula information to influence its narrative interest and to guide it towards contexts that satisfy given learning objectives. It is at this level that enters the concept of narrative cases.

### 3.3 Cases

A Case represents part of a fabula that implicitly reflects an authorial vision of an interesting sequence of events and is used in the process of "narrative inspiration"[14]. It also serves the purpose of identify points in the emerging fabula where the author viewpoint might be imposed. According to [15] the following constraints should hold:

1) A Case "…demonstrates a narrative concept. A narrative concept could for instance be 'hiding from a threat' or 'flying over an area to search for something'. The expression of a narrative concept is an implicit description of an example problem and a solution to it. (…) Of course, there can be many cases demonstrating the same narrative concept." In TARGET's context, a narrative concept can be for instance "seeding of a conflict".

2) A Case "…is context complete with regard to its narrative concept. (…) Applied to storytelling problems we define a context complete case as a case that contains all the elements necessary for the case to be viewed as 'believable' by the author of the case, regarding the narrative concept it is supposed to express, and nothing more than that". We can have elements within a case that aren't instatiated if they don't contribute for the expression of the narrative concept.

Each case should also reflect, not only an interesting sequence of events in a certain context, but also learning objectives. Each learning objective explicitly has one or more cases associated to it and these are used to "inspire" story direction. This is how we attempt to solve the problem of shaping the story landscape according to both *storyness* and *learning goals*.

The case selection mechanism is represented in Figure 4.

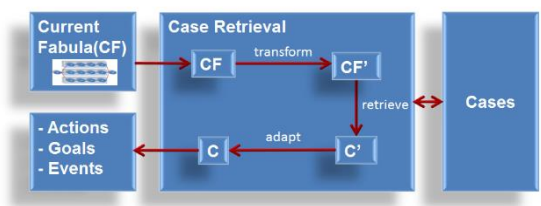

**Figure 4 - Case selection overview**

The process starts with the transformation of the current fabula to expand the space of possible applicable cases. At the time, we are performing generalization transformations based on the knowledge representation of the world domain (e.g. "Talk with a person" can be generalized to "Communicate with a person"). When a Case is retrieved we have to apply the inverse transformations previously made to adapt the case to the current context. Finally, suggested characters' actions, characters' goals or simulation events are returned based on the retrieved case.

This process identifies in the current emerging fabula, situations similar to the ones represented in the cases. When at least one Case is retrieved, the Drama Manager has the opportunity to intervene in the story and direct it. This "direction" relates to the process of "Narrative Control" and a number of issues arise when it is used with emergent narrative.

The topic of narrative control clashes with the topic of emergent narrative because we want believable autonomous characters but, at the same time, to satisfy a plot structure. The Drama Manager follows the "Drama Management properties" identified by Swartjes in his PhD thesis[16], based on Mateas & Stern's work:

1)  If autonomous characters are used, drama management is incremental.
2)  If autonomous characters are used, drama management must be opportune.
3)  If autonomous characters are used, story-level goals must be optional.

These properties reflect the facts of building on what has already emerged instead of taking the future into account (1), pursue story-level goals when opportunities occur to achieve them instead of coerce the event sequence (2) and since guidance is unreliable, abandoning story-level goals does not make the simulation fail (3). In practice, (1) is satisfied because we retrieve narrative cases based on the fabula built with the events so far, (2) is satisfied by the fact of waiting for events to occur until a case is applicable and (3) is satisfied simply because the simulation will not fail when a story-level goal is not achieved.

To achieve a target story goal, characters have to be influenced by the "Drama Manager" to do the "interesting" thing given certain situations. However, characters should also stay in character (IC) in order to maintain their believability. Directing autonomous characters is a research topic addressed in a number of works as referenced in [17, 18, 19]. According to Blumberg et al., we can see external control of characters as "weighted preferences or suggestions" that are in favor or against behaviors or actions. Swartjes argued that this form of control is ultimately "unreliable" as agents may or may not follow the suggestions but concluded that "It is, however, the price to pay for trying to control autonomous believable agents".

Szilas investigation points towards the use of this "unreliable" guidance of believable autonomous characters through the use of a "negotiation" mechanism:

"The solution consists in rethinking the relationship between the DM and IAs in terms of cooperation rather than subordination. It is obvious that both the IAs and the DM, that is both character reasoning and narrative reasoning, are concerned with high level beliefs, actions and goals. Both also should manage characters' emotions. The problem mentioned above comes from the attempt to a priori distribute some control between the two entities, in a hierarchical manner, while this control should be negotiated" [20]. Our Drama Manager influences the emerging fabula in three different ways:

1)  Character's actions;
2)  Character's goals;
3)  Simulation events;

The first one is strongly influenced by Szilas work on combining plot direction with character believability. The second and third also work with a mechanism of negotiation to maintain consistent character personalities (2) and realistic simulation (3). The next section describes the Narrative Engine module and how the cases are used to guide the emergent narrative.

## 4   TARGET's Narrative Engine

The Narrative Engine (NE) is the module that supports the generation and control of interesting stories. The high level overview of how the NE integrates with the TARGET's Platform is presented in Figure 5.

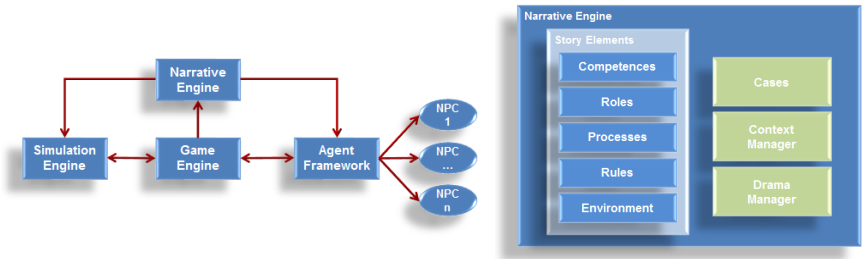

**Figure 5 – Narrative Engine (Left: NE integration, Right: NE overview)**

The GE represents the 3D story world and it is through this module that the player interacts with the game. The NPCs (Non-Player Characters) are managed and controlled by an Agent Framework (AF) and the simulation processes are handled by a Simulation Engine (SE). Given a set of Story Elements, the NE has the ability to setup and manage the storyline during a game scenario. Each Story Element will be used to instantiate either external modules or internal modules as illustrated in figure xxx.

*Competences* denote the learning goals the player wants to achieve through storytelling. To each learning goal, we associate a set of cases. This process corresponds to the instantiation of the Cases sub-module. *Roles* refer to characters' behavior as believable autonomous agents within the world simulation. Roles are used to instantiate the AF which will manage each character. *Processes* and *Rules* are the elements that represent what simulation processes will be used and the rules to which they must obey. These elements are used to instantiate the SE. *Environment* describes the space in which the story will take place. These elements are used to instantiate the GE. After the explained instantiations, the *Drama Manager* will start listening to events from the story world and starts building the emerging fabula. Whenever the *Content Manager* detects an opportunity to apply a Case (or a set of Cases) that can be applied to the current context, *narrative control* is enabled.

## 4.1 Narrative Control

Injecting narrative control according to the underling cases can be done either by triggering simulation events or suggest characters goals and actions. As explained in the previous section, narrative control shouldn't constraint simulation realism nor characters believability. Therefore, in order to manage this tradeoff a negotiation mechanism between the NE and the AF/SE was integrated in the proposed solution.

When the Drama Manager triggers an event, the SE may or may not accept that request depending on if the current simulation rules will hold or not (Figure x, left). (add example). Changing a character's goal requires the representation of Personal Values. Its according to these values that a character decides the adoption of suggested goals. Notice that influencing a character goal or a character actions has substantial different effects on believability. A goal is a general behavior while actions depend on the character's personality. The negotiation mechanism involving simulation events and agent goals is illustrated in figure 6:

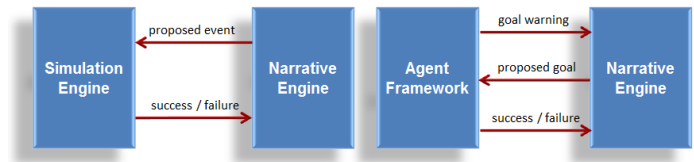

**Figure 6 - Negotiation mechanisms (left: events, right: goals)**

The negotiation mechanism that involves influencing characters actions is illustrated in figure 7:

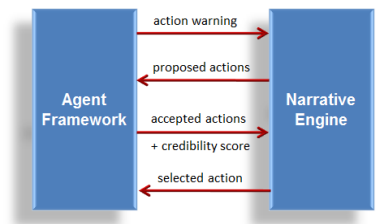

**Figure 7 - Action selection negotiation**

This mechanism was strongly influenced by Szilas work on Drama Manager and Intelligent Agent's (IA's) cooperation [20]. When the NE is warned by the AF regarding an incoming character's action, it suggests a set of meaningful actions to the AF. These actions have a correspondent narrative interest score that is calculated according to the level of resemblance between the current situation and the Cases. These set of actions are then classified by the AF according to its credibility, which reflects each action believability. The NE then confronts this information with the level of narrative interest and chooses the most appropriate action. The process of confronting believability and narrative interest is based on balanced combination of a set of thresholds.

Finally the action is sent to the AF and played by the character. Note that the set of proposed actions may not contain any "acceptable action". In that case, the character will not be influenced by the NE.

### 4.2 Example

A very simplified example is presented using the current implementation of the approach proposed in this paper. In Figure 8 is a *representation* of an authored case about "conflict seeding", applicable when teaching the "conflict management" competence. The character "Adam" is the protagonist and the character "Bob" is the antagonist.

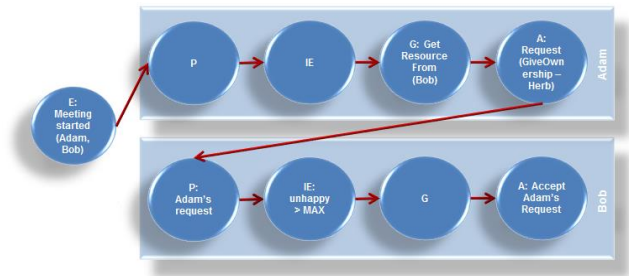

**Figure 8 - Case example.**

This case represents a story where Adam and Bob are in a meeting and Adam wants to have a human resource from Bob (named Herb). Despite of Bob being very unhappy about Adam's request the NE wants him to accept it since this situation can be the root of an intense conflict (imagine, for instance, Herb doing a poor job working for Adam because of its loyalty to Bob).

Let's consider that the emerged fabula context matches the case (as explained in section 3.3). When Bob decides to do an action the AF warns the NE about it and the "Accept Adam's Request" action (or an adaptation of it) will be suggested. If the AF complies with this suggestion, Bob will do it. At this point, authorial content on how a conflict can be seeded was successfully introduced. Note that the case can be transformed and adapted to match different similar contexts, for instance, characters can have other name and the protagonist can perform other actions instead of "GiveOwnership".

## 5 Conclusions

In this paper, we presented a narrative engine that supports emergent storytelling for the development of complex competences. In doing so, we looked at how the fabula module could be use to generate dramatic stories with human significance and learning goals. Also, we described how cases can represent situated contexts associated to the particular competences. Also, we explained how the cases could be use to build meaningful stories and drive the drama manager control of the ongoing story. The describe narrative control mechanism consist in ...